\documentclass[letter,
               ]{jacow}
\usepackage{pdfpages,multirow,ragged2e} %
\makeatletter%
	\ifboolexpr{bool{xetex}}
	 {\renewcommand{\Gin@extensions}{.pdf,%
	                    .png,.jpg,.bmp,.pict,.tif,.psd,.mac,.sga,.tga,.gif,%
	                    .eps,.ps,%
	                    }}{}
\makeatother

\ifboolexpr{bool{xetex} or bool{luatex}} 
 {}                                      
 {\usepackage[utf8]{inputenc}}           

\usepackage[USenglish]{babel}
\usepackage{amsmath}
\ifboolexpr{bool{jacowbiblatex}}%
 {%
  \addbibresource{jacow-test.bib}
  \addbibresource{biblatex-examples.bib}
 }{}
\begin{document}

\title{Next generation direct RF sampling LLRF control and monitoring system for linear accelerators 
\thanks{This work was supported by the U.S. DOE, Office of Science contract DE-AC02-76SF00515.}}

\author{C. Liu\thanks{chaoliu@slac.stanford.edu}, E. Snively, R. Herbst, K. Kim, E. A. Nanni \\ SLAC National Accelerator Laboratory, Menlo Park, California, USA \\
		}
	
\maketitle

\begin{abstract}
   The low-level RF (LLRF) systems for linear accelerating structures are typically based on heterodyne architectures. The linear accelerators normally have many RF stations and multiple RF inputs and outputs for each station, so the complexity and size of the LLRF system grows rapidly when scaling up. To meet the design goals of being compact and affordable for future accelerators, or upgrading existing ones, we have developed and characterized the next generation LLRF (NG-LLRF) platform based on the RF system-on-chip (RFSoC) for S-band and C-band accelerating structures. The integrated RF data converters in RFSoC sample and generate the RF signals directly without any analogue mixing circuits, which significantly simplified the architecture compared with the conventional LLRF systems. We have performed high-power tests for the NG-LLRF with the S-band accelerating structure in the Next Linear Collider Test Accelerator (NLCTA) test facility at SLAC National Accelerator Laboratory and a C-band structure prototyped for Cool Cooper Collider (CCC). The NG-LLRF platform demonstrated pulse-to-pulse fluctuation levels considerably better than the requirements of the targeted applications and high precision and flexibility in generating and measuring the RF pulses. In this paper, the characterization results of the platform with different system architectures will be summarized and a selection of high-power test results of the NG-LLRF will be presented and analyzed.
\end{abstract}

\section{Introduction}

The conventional low-level RF (LLRF) systems of particle accelerators are typically implemented with heterodyne-based architectures. The RF signals are up and down mixed from the baseband with analog RF mixers. There are many RF input and output channels in large accelerators, so the size and cost of analog circuits of the LLRF system could become substantial. With the latest direct RF sampling technology of the RF system-on-chip (RFSoC), we have designed and implemented the next-generation LLRF (NG-LLRF) system for linear accelerators (LINACs) \cite{liu2024next}, including the future collider concept, the Cool Cooper Collider (C\(^3\))\cite{emilio} and the upgrade solution for the S-band accelerator test station at the Next Linear Collider Test Accelerator (NLCTA)\cite{liu2025ipacS}, which have stringent RF performance requirements for LLRF systems. 

RFSoC technology has been applied to a range of high-energy physics and astrophysics instruments with RF frequencies up to 6 GHz and we have performed RF performance characterizations with different RFSoC devices specifically configured for different applications \cite{liu2021characterizing ,henderson2022advanced,liuRA,liuRF,liu2023higher,liuSQ}. The development of NG-LLRF at SLAC begins with a C-band linear accelerator(LINAC). The prototype has demonstrated low amplitude and phase fluctuation levels for C-band applications \cite{liu:ipac2024-mocn2,liu2025high,liu2025accel} and a selection of the evaluation results will be summarized in this paper. There are many S-band RF stations around SLAC, including the S-band test stations in test laboratories, the S-band station in NLCTA, and the RF guns and approximately 200 klystrons of LCLS \cite{lcls} and FACET. The NG-LLRF could be an excellent candidate for upgrading the LLRF systems for those RF station. To start the exercise, we have configured the NG-LLRF for the S-band test station at NLCTA and performed some initial performance evaluation \cite{liu2025ipacS}. In this paper, the test configuration used for the initial high-power test results of the NG-LLRF with the S-band station at NLCTA will be introduced and some test results will be discussed.      

\section{NG-LLRF for C-band LINACs}

\subsection{NG-LLRF System Architecture} 

\begin{figure*}[!htb]
   \centering
   \includegraphics*[width=\textwidth]{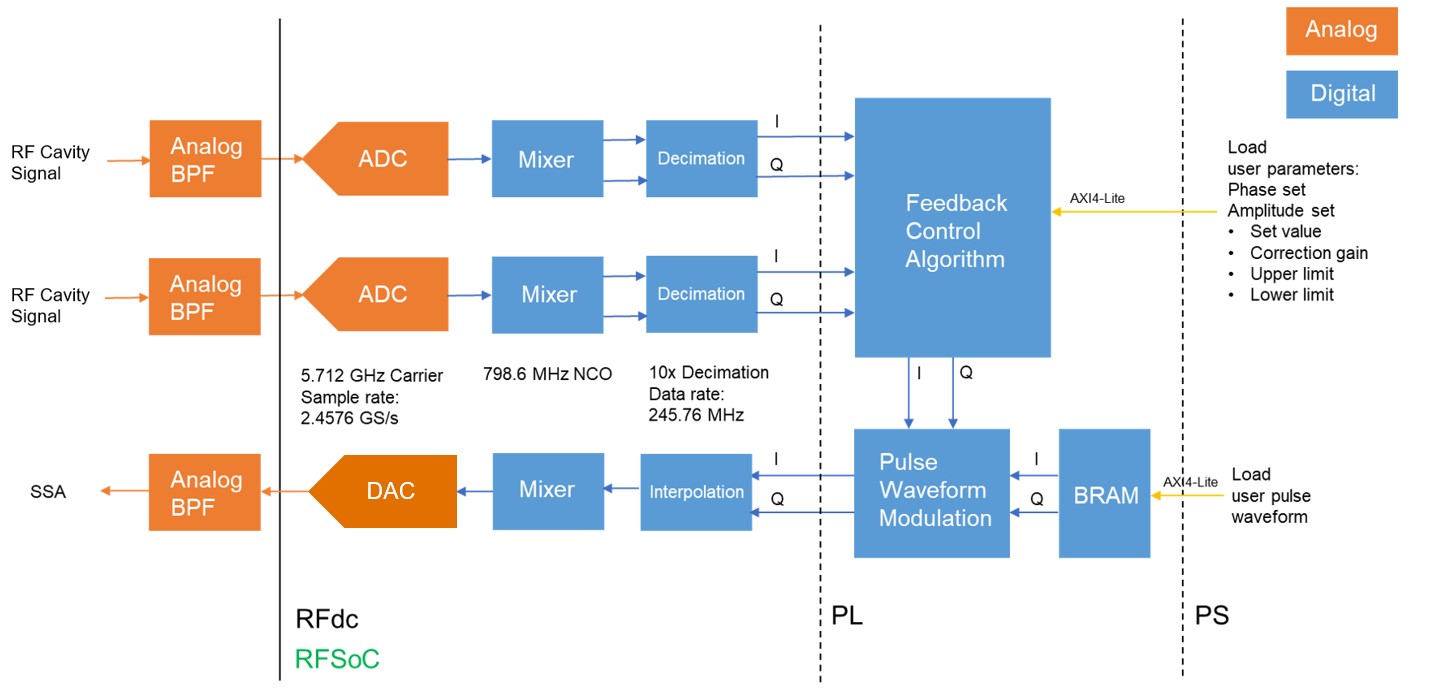}
   \caption{The block diagram of the NG-LLRF control circuit for C-band LINAC. All the components or firmware blocks on the right side of the solid line are integrated or implemented in RFSoC\cite{liu2024next}.}
   \label{fig:f1}
\end{figure*}

Figure \ref{fig:f1} shows the block diagram of the NG-LLRF with two rf input channels and a single rf output channel for a single set of accelerating structure cavities. However,the NG-LLRF system can have up to 16 rf input channels and 16 rf output channels, which can be customized per application. 

The center frequency of the C-band NG-LLRF is around 5.712 GHz and the bandwidth is 245.76~MHz. The ADCs direct sample the RF signals with sampling rate of 2.4576 giga samples per second (GSPS) at higher order Nyqusit zone. The digital samples are digitally down-mixed  in a harden IP implemented in RFSoC. The baseband output of the mixer in in-phase (I) and quadrature (Q) format is then decimated and filtered. The feedback control algorithm block in the firmware reads IQ components of the down-converted cavity signal and calculates a new set of IQ values based on user input parameters. The updated IQ is modulated with the user-defined baseband pulse. The modulated pulse is then interpolated and digitally up-converted with a hardened IP implemented in the RFSoC. The updated RF pulse is generated by the integrated DAC in RFSoC at sampling rate of 5.89824 GSPS and the RF frequency is in the second Nyquist zone.

\begin{figure}[!htb]
   \centering
   \includegraphics*[width=1\columnwidth]{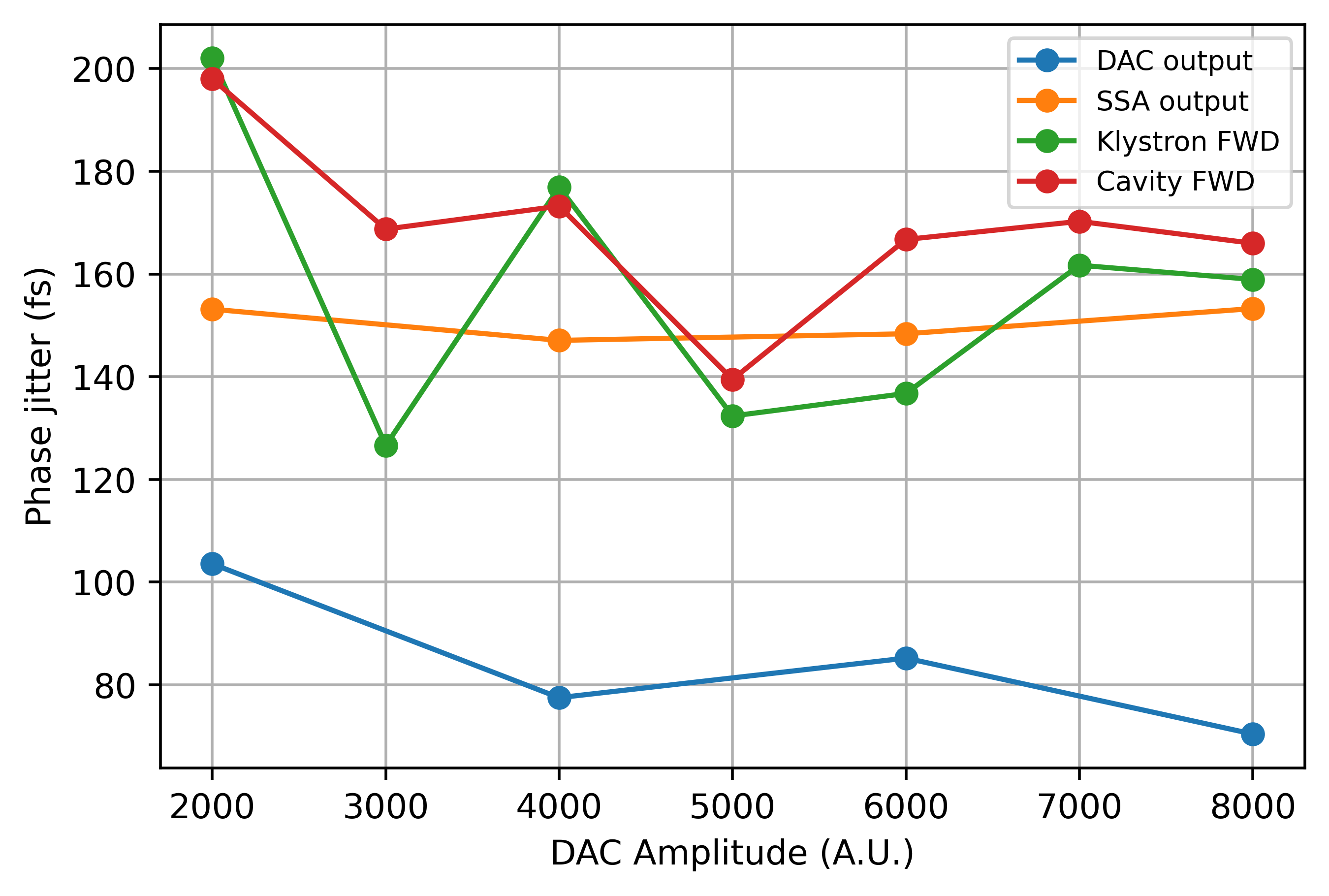}
   \caption{The phase jitter levels at different stage of the driving circuit at different power levels.\cite{liu2025high}}
   \label{fig:f2}
\end{figure}

\begin{figure}[!htb]
   \centering
   \includegraphics*[width=1\columnwidth]{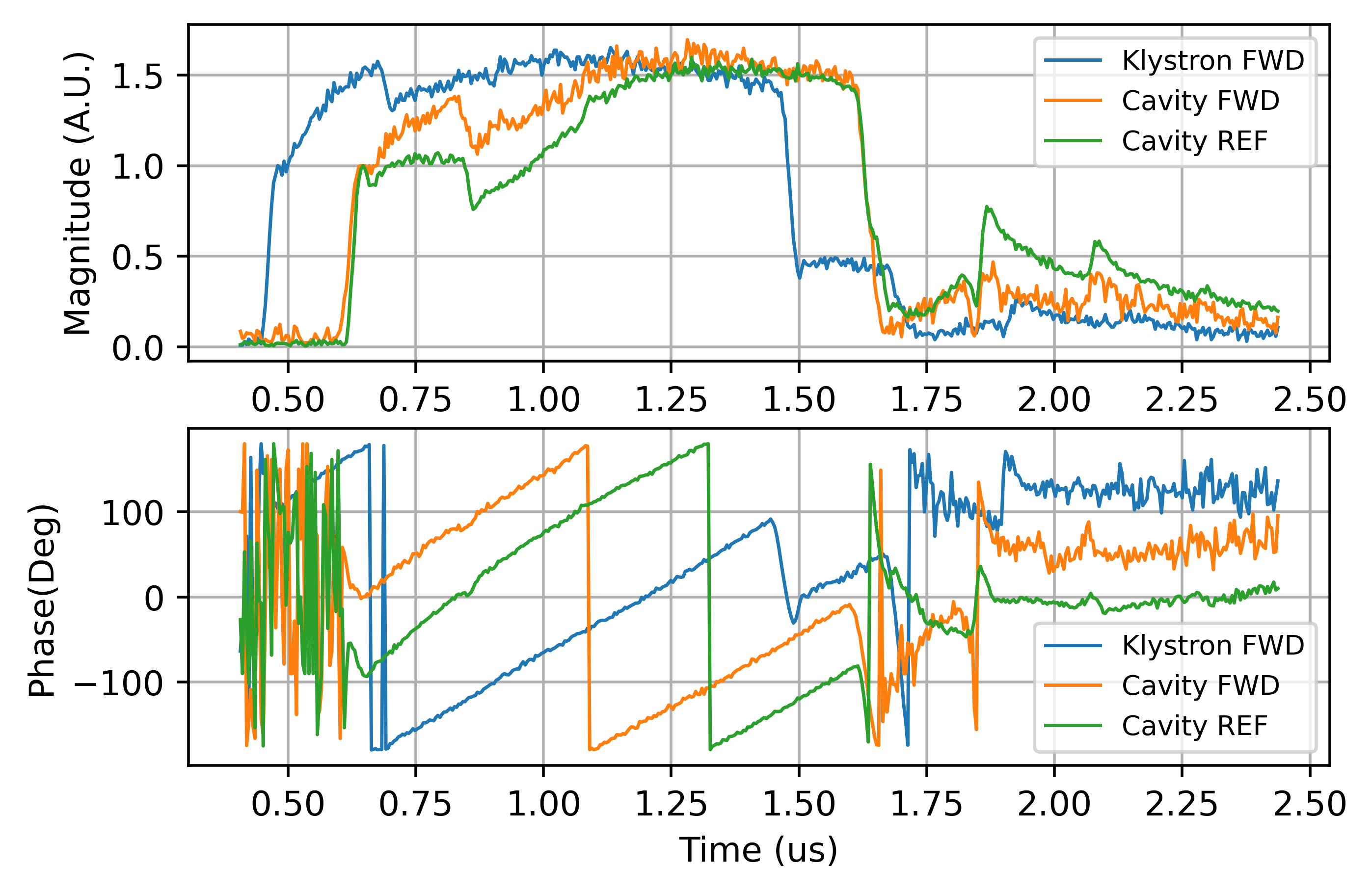}
   \caption{The magnitude and phase of the base-band pulses measured with a 1 \(\mu\)s pulse with a 360\textdegree linear phase ramp at the peak power of 5.17 MW.\cite{liu2025ipacC}}
   \label{fig:f3}
\end{figure}

\begin{figure*}[!htb]
   \centering
   \includegraphics*[width=\textwidth]{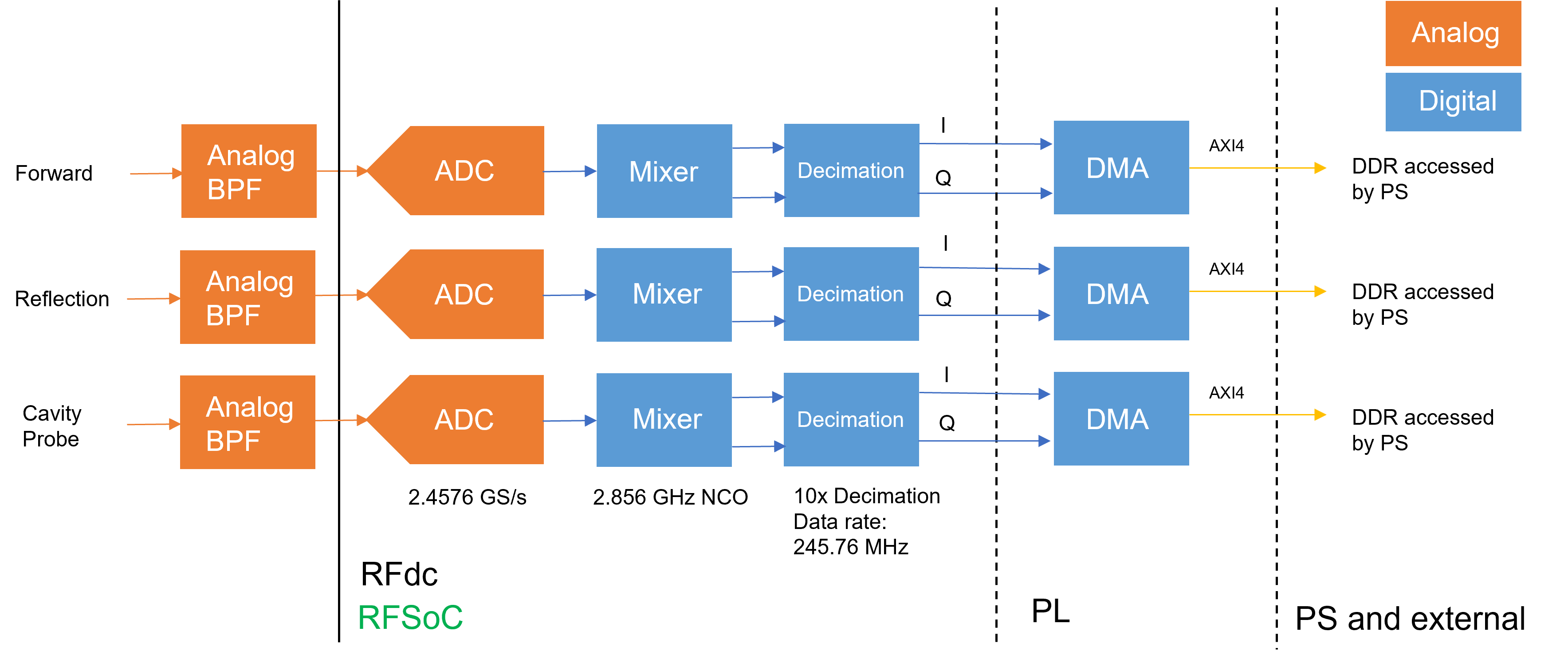}
   \caption{The block diagram of the NG-LLRF configured for S-band structure measurements.}
   \label{fig:f4}
\end{figure*}

\subsection{Pulse-to-pulse Stability with Highe-power Test} 

Pulse-to-pulse fluctuation level is critical for an LLRF system. We measured phase jitter within 60 consecutive RF pulses at four different stages with appropriate attenuation, the DAC RF signal directly measured by ADC, the SSA output coupler, the klystron forward and the cavity forward. The measurements was made at different power levels, which is noted as the DAC amplitude used to modulate with the RF signal. With DAC amplitude from 2000 to 8000, the peak RF power injected into the structure ranges from about 4.2 MW to 16.45 MW. 

The phase jitter with direct loopback is around 80 ns when the DAC amplitude s 4000 or higher. As the orange trace in Figure \ref{fig:f2} shows, the SSA added around 70 fs of additional phase jitter to the rf pulse compared with the direct loop-back test. The phase jitter levels of klystron forward fluctuate around the SSA jitter level, which is approximately 150 fs. The phase jitter added by the field coupling from the klystron to the accelerating structure is in the range of 10 to 40 fs. With 16.45 MW delivered to the accelerating structure, the added phase jitter by the SSA, klystron and waveguide are around 82.8 , 5.6 and 7.0 fs respectively, which totals the final phase jitter at cavity forward coupler to be 166 fs. With a feedback control, the 150 fs phase jitter requirement of C\(^3\) is highly achievable with the NG-LLRF platform. The development of the real-time feedback control loop is still in progress, and more test results will be published with a full implementation of the NG-LLRF system.

\subsection{Arbitrary Pulse Shaping with NG-LLRF in C-band} 

Since modulation and demodulation of the RF pulse are fully implemented in the digital domain, the RF pulse can be flexibly shaped. Figure \ref{fig:f3} shows the baseband pulses captured at different stages when the test stand was driven by RF pulses with a linear phase ramp. The klystron forward and cavity forward signals show that the phase ramp has been successfully introduced. However, the RF power injected into the structure is fully reflected, as the phase ramp is equivalent to drive the structure off resonance. That demonstrated the phase modulation and measurement precision of the NG-LLRF platform, which are critical techniques for pulse compressors and beam loading compensation.

\section{NG-LLRF for S-band LINACs}

\subsection{High-power Test Setup} 

The NG-LLRF reconfigured for an S-band frequency of around 2.856 GH was integrated with the S-band station at NLCTA. As Figure \ref{fig:f4} shows, the integration was started using the NG-LLRF to measure the RF signals at different stages of the accelerating structure. The RF signals from the accelerating structure are digitized by the ADCs integrated in the RFSoC at 2.4576 GSPS and then down mixed digitally with the 2.856 GHz signal generated by the numerically controlled oscillator (NCO). The digital mixer converts the signal in real format to in-phase (I) and quadrature (Q) format. The IQ components are decimated by a factor of 10, so the RF bandwidth is 245.76 MHz in this case. The decimation can be customized to other values to achieve a different bandwidth. The NCO frequency can be configured in software for other down conversion frequencies.

\begin{figure}[!htb]
   \centering
   \includegraphics*[width=1\columnwidth,angle=-90]{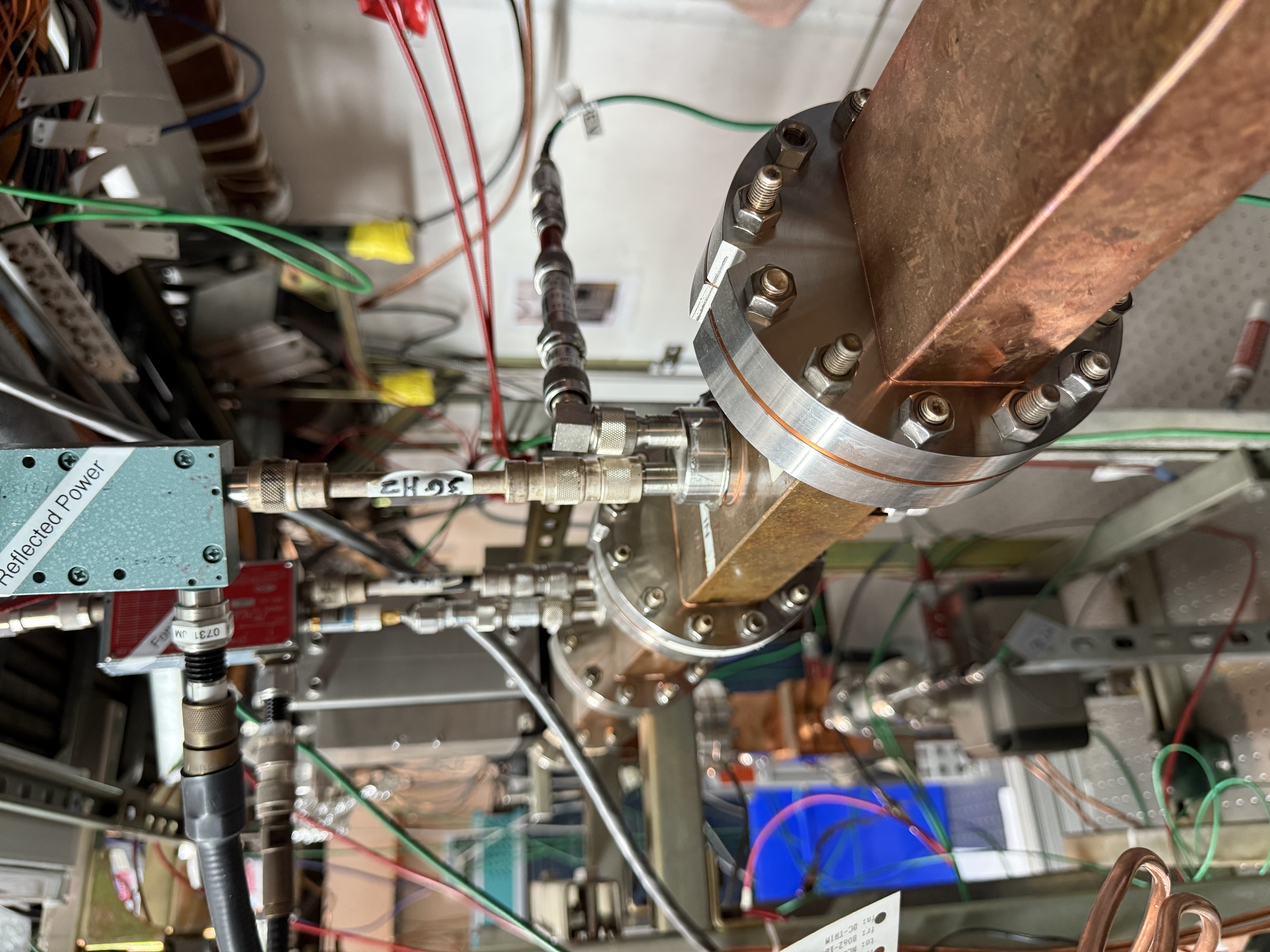}
   \caption{The forward and reflection couplers of S-band test structure in NLCTA bunker.}
   \label{fig:f5}
\end{figure}

\begin{figure}[!htb]
   \centering
   \includegraphics*[width=1\columnwidth,angle=-90]{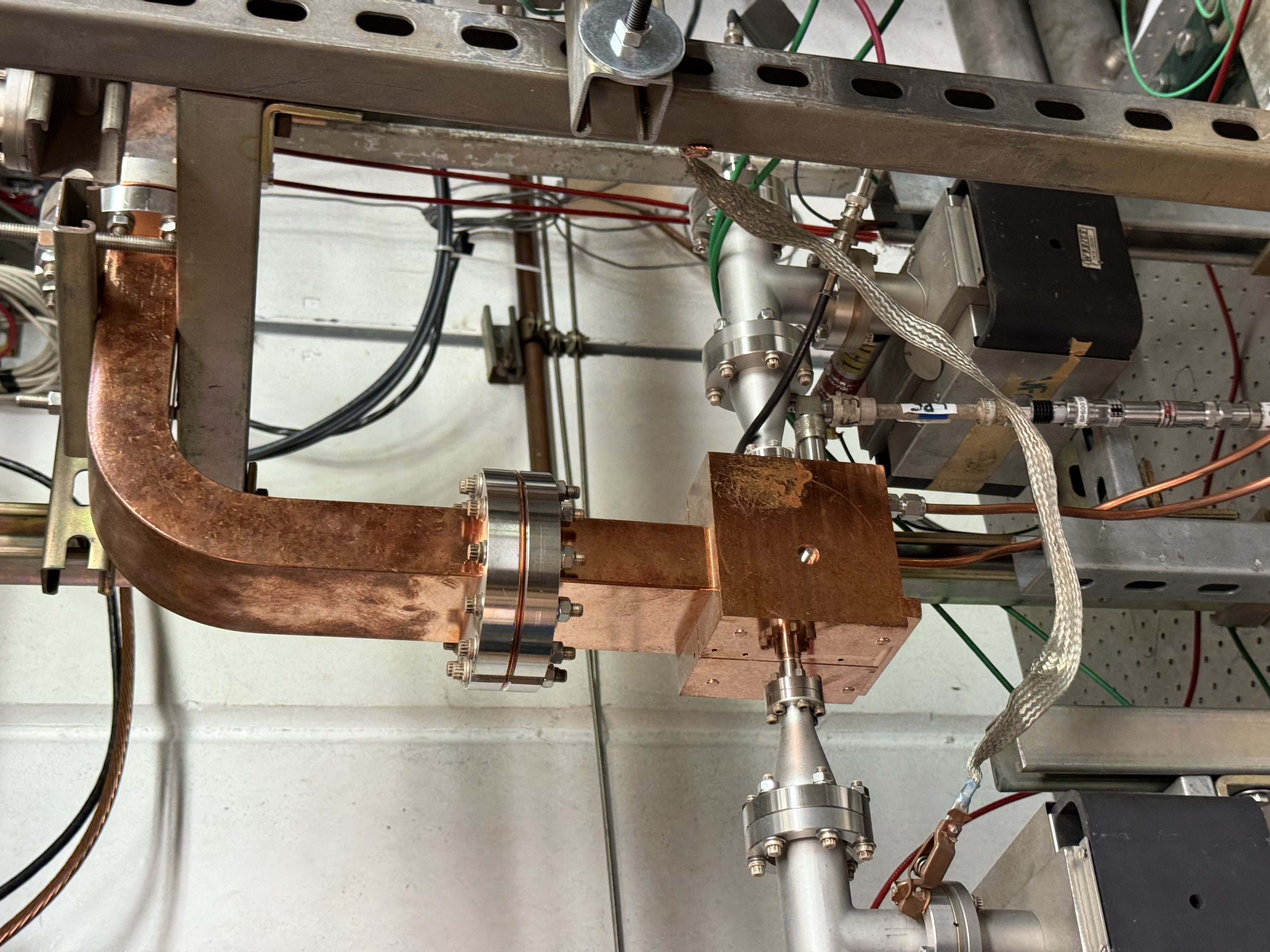}
   \caption{The cavity probe coupler of S-band test structure in NLCTA bunker. }
   \label{fig:f6}
\end{figure}

Figure \ref{fig:f5} shows the couplers that sample forward and reflected power. Figure \ref{fig:f6} shows the coupler that samples the cavity probe power level. Power attenuators were inserted for the three measurement points to limit the power levels at the RFSoC ADC input not greater than 0 dBm.

\subsection{High-power Test Results} 

In the initial test, we used a Hittite function generator to generate a continuous wave (CW) RF signal around 2.856 GHz. RF pulses were generated by an R\&K power amplifier with the triggering and gate signal supplied to it. The duration of the RF pulses in this test is approximately 2 \(\mu\)s The test was performed with three output power levels of the function generator, which are 3 dBm, -3 dBm and -9 dBm. We did not perform power mapping in this case, as we are aiming to see if we are able to capture the RF pulse with NG-LLRF. More careful tests with full power mapping will be performed at a later stage of system integration. 

\begin{figure}[!htb]
   \centering
   \includegraphics*[width=1\columnwidth]{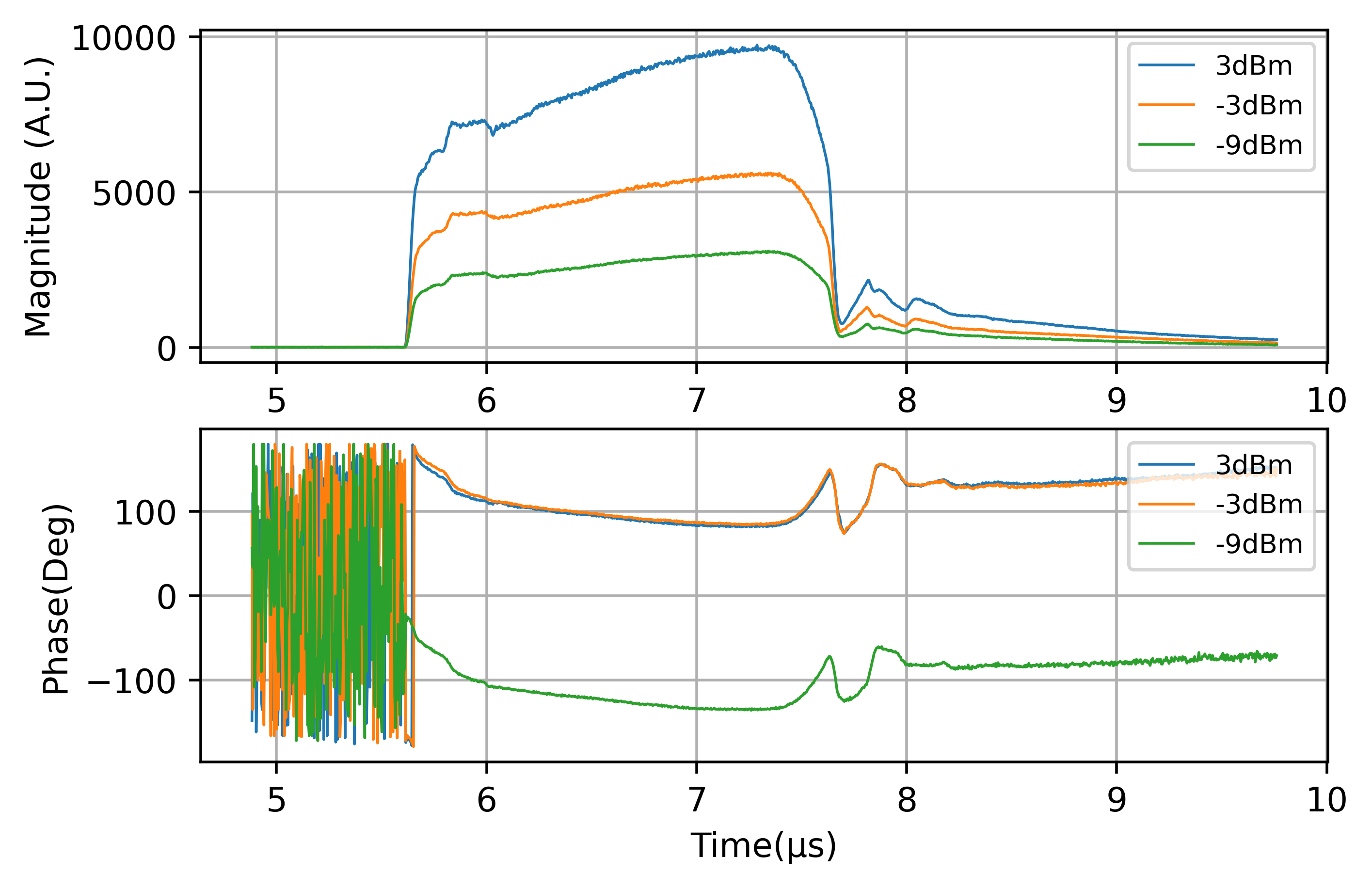}
   \caption{he RF pulse captured from the forward coupler of the S-band structure in NLTCA test facility. The RF pulses forwarded to the structure at three different drive power levels have been captured}
   \label{fig:f7}
\end{figure}

Figure \ref{fig:f7} shows the basedband pulses captured at the cavity forward couplers at three different power levels. The trends of magnitude and phase are consistent for measurements at different power levels, which demonstrates the measurement precision of the NG-LLRF in S-band. The rise time of the forward power is roughly 0.4 \(\mu\)s, which depends mainly on rise time of the klystron. After the rise time, the forward power continues to increase with a mellow slop until the RF is off. For a single bunch system, the slope does not affect the performance as long as it is consistent pulse to pulse.

\begin{figure}[!htb]
   \centering
   \includegraphics*[width=1\columnwidth]{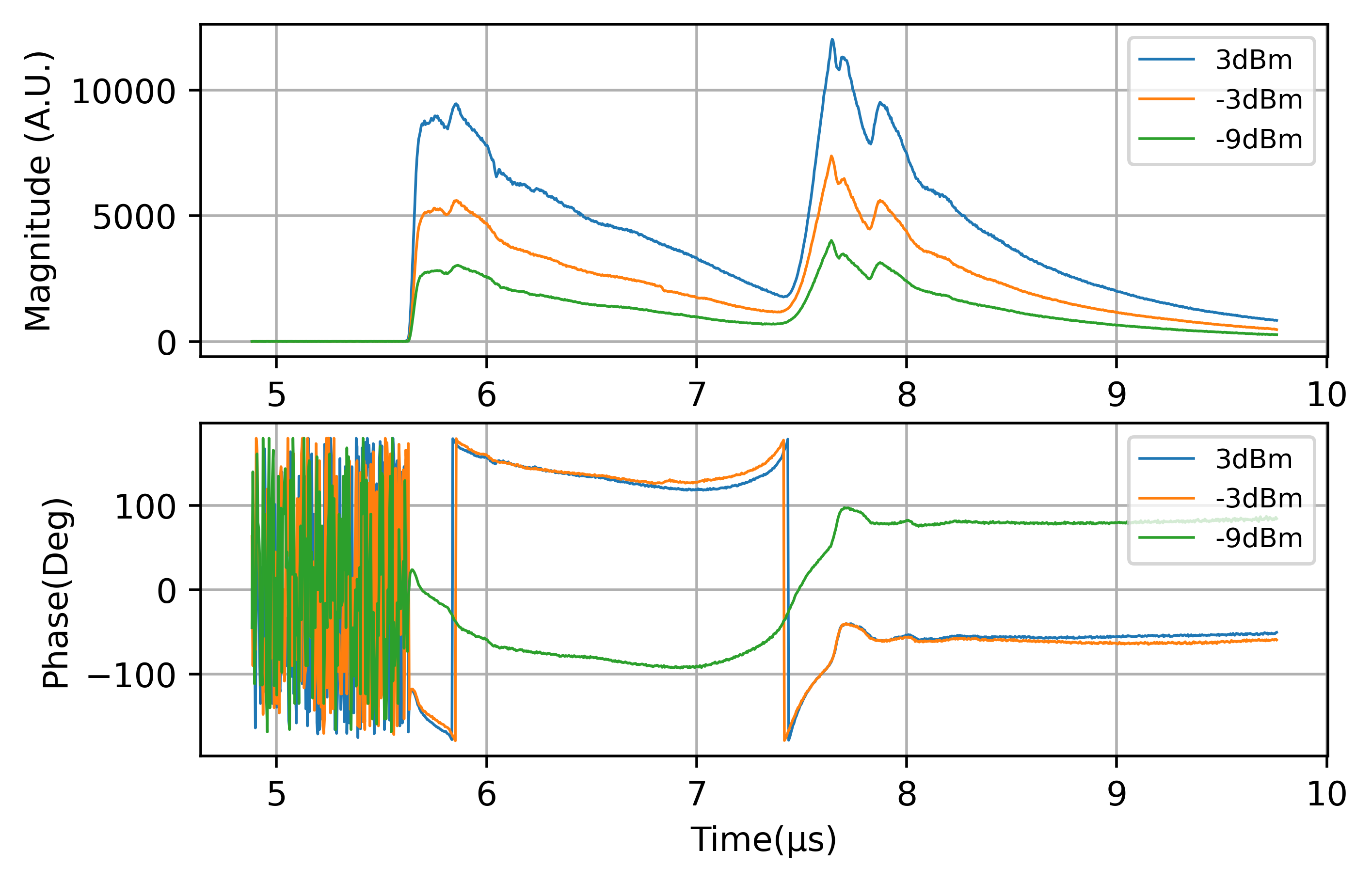}
   \caption{The RF pulses captured from the forward coupler of the S-band structure in NLTCA test facility. The RF pulses forwarded to the structure at three different drive power levels have been captured.}
   \label{fig:f8}
\end{figure}

Figure \ref{fig:f8} shows the basedband pulses captured at the reflection couplers at three different power levels. The reflection signals reveal the full field filling process in the accelerating structure. After the rise time of the klystron, the reflections signal decreases linearly as the field filling the cavity. When the RF is switched off, the power is reflected back to the waveguide and decays as the RF power dissipated as heat. The system was designed to be over coupled, so the second peak of the reflection signal is higher than the first one.

\begin{figure}[!htb]
   \centering
   \includegraphics*[width=1\columnwidth]{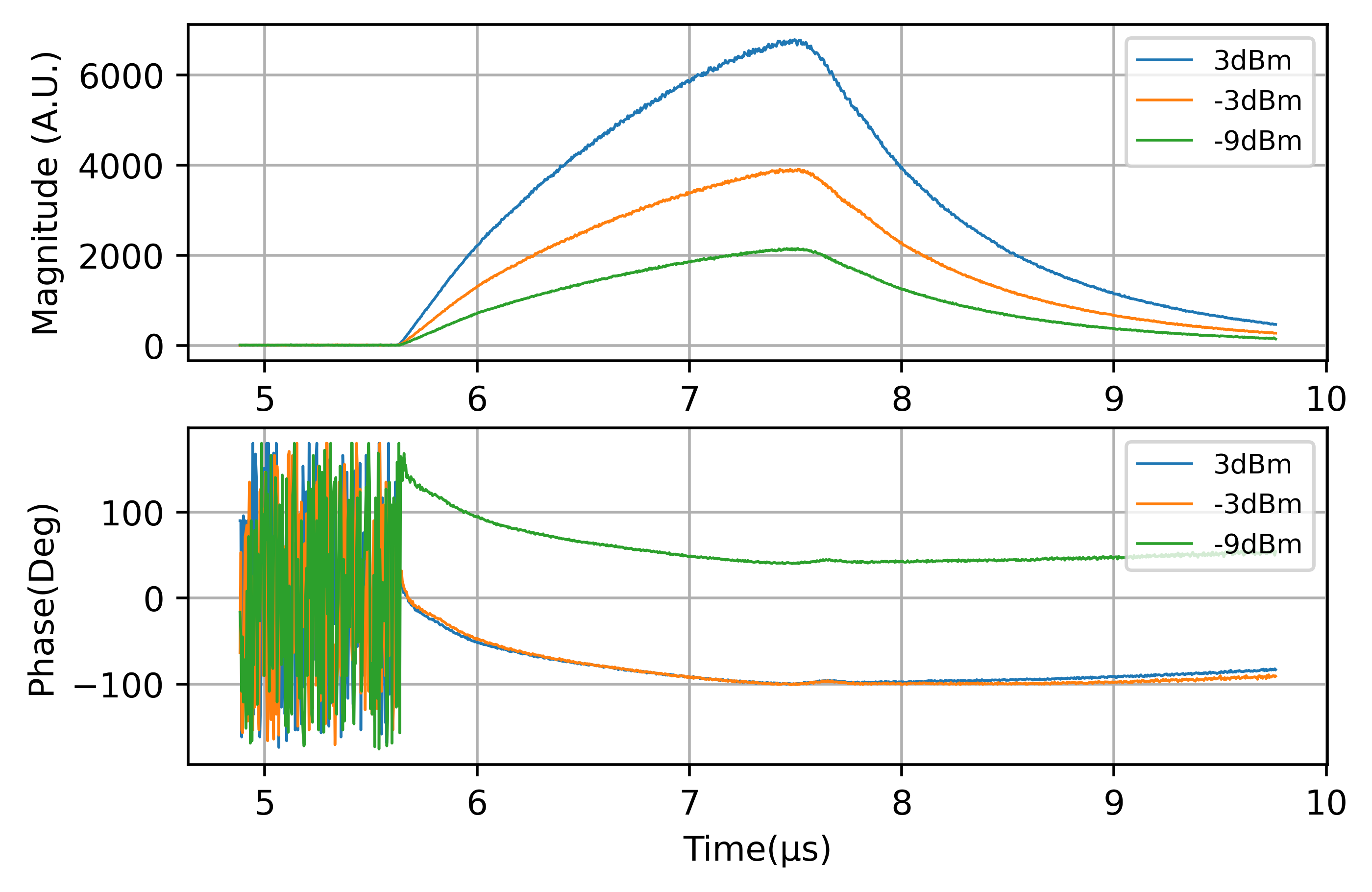}
   \caption{The RF pulses captured from the cavity probe of the S-band structure in NLTCA test facility. The RF power injected to the structure is coupled to the probe and the RF pulse from the probe at three different drive power levels have been captured.}
   \label{fig:f9}
\end{figure}

Figure \ref{fig:f9} shows the basedband pulses captured at the cavity probe couplers at three different power levels, which directly reveal the field filling process. The power in the cavity increases over the full RF pulse and decays gradually after the RF is switched off.

\subsection{NG-LLRF Enclosure} 

\begin{figure}[!htb]
   \centering
   \includegraphics*[width=1\columnwidth]{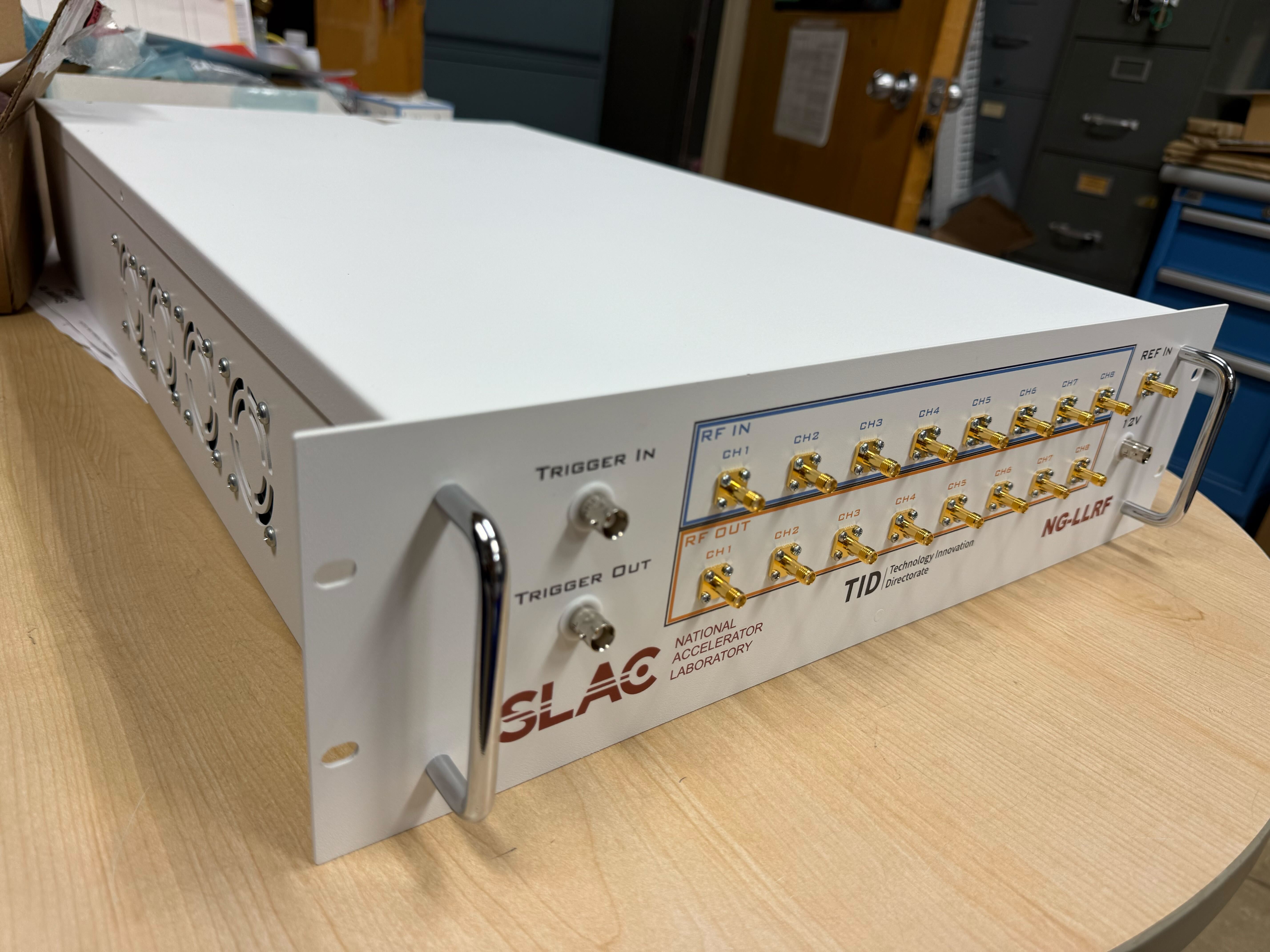}
   \caption{The rack mount NG-LLRF chassis with a ZCU208 RFSoC board.}
   \label{fig:f10}
\end{figure}

We have designed a NG-LLRF chassis, which is shown in Figure \ref{fig:f10}. Figure \ref{fig:f10} shows the front panel of the chassis, which has 8 RF input channels and 8 RF output channels. The RF input and output channels are connected to the integrated ADCs and DACs in the RFSoC. Depending on the application, the RF channels can be individually selected and configured. There are both trigger input and trigger output ports on the front pane, which means the NG-LLRF can be either the slave for a master trigger source or the master trigger for the system. By connecting a reference to the "REF IN" port, the NG-LLRF can be locked to an external reference signal. By default, the system will be locked to a 10 MHz reference. For different reference at other frequencies or other specific clocking requirements, the PLL chip on the RFSoC can be reprogrammed by loading a new register file on to an SD card that boots up the RFSoC system. The 12V power supply port is reserved for additional RF amplifier or other active parts need to be added to a system that requires a separate supply with higher stability.

In the back plane of the chassis, we have an RJ45 GbE for remote update of firmware and software and data transfer at lower rate. There are two SFP 10 GbE ports for data transfer at higher rates and timing distribution. The enclosure is power from a single cord from the main and there is a power supply module that generates the required DC supplies required by the system. The RFSoC operation can dissipate excess heat, and fans are on both sides of the chassis for ventilation.

\subsection{Future Development Plan} 

We are in progress in integrating the NG-LLRF with the NLCTA S-band station. The next step is to interface with the clock and triggering system of the S-band station with NG-LLRF and characterize the RF performance of the NG-LLRF. Then we would use the integrated DAC in RFSoC to generate the RF pulse that drives the S-band station. When the full feedback control scheme is implemented in firmware, the NG-LLRF will replace the existing LLRF system.

\section{Summary}

 The preliminary results of the high-power test verified that NG-LLRF can measure the RF pulses from the accelerating structure with high accuracy in both C-band and S-band. The NG-LLRF demonstrated high pulse-to-pulse stability and high flexibility in pulse shaping in the C-band. More performance characterization results will be published with NG-LLRF fully integrated to the S-band station at NLCTA. The full realtime feedback control firmware is still under development. The software, firmware and hardware architecture has been designed in a modularized approach to maximize the adaptability of the platform. 
%
%
\ifboolexpr{bool{jacowbiblatex}}%
	{\printbibliography}%
	{%
	

} 
%
%


\end{document}